     \documentstyle[aps,prl,twocolumn,epsf]{revtex}

\newcommand{\fpi}{\mbox{$F_\pi$}}
\newcommand{\qsq}{\mbox{$Q^2$}}
\newcommand{\sigl}{\mbox{$\sigma_{\mathrm{L}}$}}
\newcommand{\sigt}{\mbox{$\sigma_{\mathrm{T}}$}}
\newcommand{\siglt}{\mbox{$\sigma_{\mathrm{LT}}$}}
\newcommand{\sigtt}{\mbox{$\sigma_{\mathrm{TT}}$}}
\newcommand{\eps}{\mbox{$\epsilon$}}

\newcommand{\gevsq}{\mbox{$\text{(GeV}/c\text{)}^2$}}

\begin{document}
\draft
\title{Measurement of the charged pion electromagnetic form factor}
\author{
J. Volmer$^{11,21}$, 
D. Abbott$^{19}$, 
H. Anklin$^{4,19}$,  
C. Armstrong$^{2}$, 
J. Arrington$^{1}$,
K. Assamagan$^{5}$,
S. Avery$^{5}$,
O.K. Baker$^{5,19}$,
H.P. Blok$^{11,21}$,
C. Bochna$^{6}$, 
E.J. Brash$^{16}$, 
H. Breuer$^{8}$, 
N. Chant$^{8}$,
J. Dunne$^{19}$,
T. Eden$^{12,19}$, 
R. Ent$^{19}$,
D. Gaskell$^{14}$,  
R. Gilman$^{17,19}$, 
K. Gustafsson$^{8}$, 
W. Hinton$^{5}$,
G.M. Huber$^{16}$,
H. Jackson$^{1}$, 
M.K. Jones$^{2}$,
C. Keppel$^{5,19}$, 
P.H. Kim$^{7}$, 
W. Kim$^{7}$, 
A. Klein$^{13}$, 
D. Koltenuk$^{15}$,
M. Liang$^{19}$, 
G.J. Lolos$^{16}$, 
A. Lung$^{19}$,
D.J. Mack$^{19}$, 
D. McKee$^{10}$,
D. Meekins$^{2}$, 
J. Mitchell$^{19}$,
H. Mkrtchyan$^{22}$,
B. Mueller$^{1}$,
G. Niculescu$^{5}$,
I. Niculescu$^{5}$,
D. Pitz$^{18}$,
D. Potterveld$^{1}$, 
L.M. Qin$^{13}$,
J. Reinhold$^{1}$, 
I.K. Shin$^{7}$, 
S. Stepanyan$^{22}$, 
V. Tadevosyan$^{22}$, 
L.G. Tang$^{5,19}$,
R.L.J. van der Meer$^{16,19}$, 
K. Vansyoc$^{13}$,
D. Van Westrum$^{3}$, 
W. Vulcan$^{19}$, 
S. Wood$^{19}$, 
C. Yan$^{19}$, 
W.-X. Zhao$^{9}$, 
B. Zihlmann$^{19,20}$\\
(The Jefferson Lab \fpi\ Collaboration)
}
\address{
$^{1}$Argonne National Laboratory, Argonne, Illinois 60439\\
$^{2}$College of William and Mary, Williamsburg, Virginia 23187\\
$^{3}$University of Colorado, Boulder, Colorado 76543\\
$^{4}$Florida International University, Miami, Florida 33119\\
$^{5}$Hampton University, Hampton, Virginia 23668\\
$^{6}$University of Illinois, Champaign, Illinois 61801\\
$^{7}$Kyungpook National University, Taegu, Korea\\
$^{8}$University of Maryland, College Park, Maryland 20742\\
$^{9}$M.I.T.--Laboratory for Nuclear Sciences and Department of Physics, 
      Cambridge, Massachussetts 02139\\
$^{10}$New Mexico State University, Las Cruces, New Mexico 88003-8001\\
$^{11}$NIKHEF, Postbus 41882, NL-1009 DB Amsterdam, The Netherlands\\
$^{12}$Norfolk State University, Norfolk, Virginia 23504\\
$^{13}$Old Dominion University, Norfolk, Virginia 23529\\
$^{14}$Oregon State University, Corvallis, Oregon 97331\\
$^{15}$University of Pennsylvania, Philadelphia, Pennsylvania 19104\\
$^{16}$University of Regina, Regina, Saskatchewan S4S-0A2, Canada\\
$^{17}$Rutgers University, Piscataway, New Jersey 08855\\
$^{18}$DAPNIA/SPhN, CEA/Saclay, F-91191 Gif-sur-Yvette, France\\
$^{19}$Physics Division, TJNAF, Newport News, Virginia 23606\\
$^{20}$University of Virginia, Charlottesville, Virginia 22901\\
$^{21}$Faculteit Natuur- en Sterrenkunde, Vrije Universiteit, NL-1081 HV 
        Amsterdam, The Netherlands\\
$^{22}$Yerevan Physics Institute, 375036 Yerevan, Armenia\\
}

\date{\today}
 
\maketitle

\begin{abstract}
Separated longitudinal and transverse structure functions for the reaction  
$^1$H$(e,e'\pi^+)n$ were measured in the momentum transfer region 
\qsq\ = 0.6 - 1.6 \gevsq\ at a value of the invariant mass $W = 1.95$ GeV. New 
values for the pion charge form factor were extracted from the longitudinal 
cross section by using a recently developed Regge model. The results 
indicate that the pion form factor in this region is  larger than
previously assumed and
 is consistent with a monopole parameterization fitted to very
low \qsq\ elastic data.
\end{abstract}


\narrowtext


The pion occupies an important place in the study of the quark-gluon structure 
of hadrons. This is exemplified by the many calculations that treat the pion as
one of their prime examples \cite{lis92} --\cite{mar98}.
One of the reasons is that the 
valence structure of the pion, being $\langle q\bar{q}\rangle$, is relatively 
simple.  Hence it is expected that the value of the four-momentum transfer 
squared \qsq , down to which a pQCD approach to the pion structure can be 
applied, is lower than e.g. for the nucleon.  Furthermore, the asymptotic 
normalization of the pion wave function, in contrast to that of the nucleon, is
known from the pion decay.

The charge form factor of the pion, $F_\pi(Q^2)$, is an essential element of 
the structure of the pion. Its behaviour at very low values of \qsq , which is 
determined by the charge radius of the pion, has been determined up to 
\qsq =0.28 \gevsq\ from scattering high-energy pions from atomic electrons 
\cite{ame86}. For the determination of the pion form factor at higher values of
\qsq\ one has to use high-energy electroproduction of pions on a nucleon, i.e.,
employ the $^1$H$(e,e'\pi^+)n$ reaction. For selected kinematical conditions 
this process can be described as quasi-elastic scattering of the electron from 
a virtual pion in the proton. In the t-pole approximation the longitudinal 
cross section \sigl\ is proportional to the square of the pion form factor.
 In this way the pion form factor has been studied for \qsq\ values from
0.4 to 9.8 \gevsq\ at CEA/Cornell \cite{beb78} and for \qsq\ = 0.7 \gevsq\ 
at DESY \cite{bra77}.
 In the DESY experiment a longitudinal/transverse (L/T) separation
was performed by taking data at two values of the electron energy. In the
experiments done at CEA/Cornell this was done in a few cases only, and even
then the resulting uncertainties in \sigl\ were so large that the L/T
separated data were not used. Instead for the actual 
determination of the pion form factor \sigl\ was calculated
by subtracting from the measured (differential) cross section a \sigt\ that
was assumed to be proportional to the total virtual photon cross section.
No uncertainty in \sigt\ was included in this subtraction.
This means that  existing values of \fpi\ above \qsq\ = 0.7 \gevsq\
are not based on L/T separated cross sections. 
This, together with the already relatively large statistical 
(and systematic) uncertainties of those data, precludes a meaningful comparison
with theoretical calculations in that region.

Because of the excellent properties of the electron beam and experimental setup
at {CEBAF} it is now possible to determine L/T separated cross sections with 
high accuracy and thus to study the pion form factor in the regime of
\qsq\ = 0.5 - 3.0 \gevsq . Using the High Momentum Spectrometer and the Short
Orbit Spectrometer of Hall C and 
electron energies between 2.4 and 4.0 GeV, data for the reaction 
$^1$H$(e,e'\pi^+)n$ were taken for central values of \qsq\ of 0.6, 0.75, 
1.0 and 1.6 \gevsq , at a central value of the invariant mass $W$ of 1.95 GeV.

The cross section for pion electroproduction can be written as
\begin{equation}
\label{eq:sigma1}
\frac{d^3 \sigma}{dE' d\Omega_{e'} d\Omega_\pi} = \Gamma_V
\frac{d^2 \sigma}{dt d\phi},
\end{equation}
where $\Gamma_V$ is the virtual photon flux factor, $\phi$ is the azimuthal 
angle of the outgoing pion with respect to the electron scattering plane and 
$t$ is the Mandelstam variable $t=(p_\pi-q)^2$. The two-fold differential cross
section can be written as
\begin{eqnarray}
\label{eq:sepsig}
2\pi \frac{d^2 \sigma}{dt d\phi} & = & 
   \epsilon \hspace{0.5mm} \frac{d\sigma_{\mathrm{L}}}{dt} +
   \frac{d\sigma_{\mathrm{T}}}{dt} + \sqrt{2\epsilon (\epsilon +1)}
   \hspace{1mm}\frac{d\sigma_{\mathrm{LT}}}{dt}
   \cos{\phi}  \nonumber \\
   & & + \epsilon \hspace{0.5mm}
   \frac{d\sigma_{\mathrm{TT}}}{dt} \hspace{0.5mm} \cos{2 \phi} ,
 \end{eqnarray}
 where $\epsilon$ is the virtual-photon polarization parameter.
The cross sections $\sigma_X \equiv \frac{d\sigma_{\mathrm{X}}}{dt}$ depend on
$W$, \qsq\ and $t$.
The longitudinal cross section \sigl\ is dominated by the t-pole
term, which contains \fpi . The $\phi$ acceptance of the experiment allowed the
interference terms \siglt\ and \sigtt\ to be determined. Since data were taken 
at two energies at every \qsq , \sigl\ could be separated from \sigt\ by means 
of a Rosenbluth separation. 

The analysis of the experimental data included the following \cite{thesis}.
Electron identification in the Short Orbit Spectrometer was done by using the 
combination of lead glass calorimeter and gas Cerenkov containing Freon-12 at 
atmospheric pressure. Pion identification in the High Momentum Spectrometer was
largely done using time of flight between two scintillating hodoscope arrays. 
A small contamination by real electron-proton coincidences at the highest 
\qsq\ setting was removed by a single beam-burst cut on $e-\pi^{+}$ coincidence
time.  Then \qsq , $W$, $t$, and the mass of the undetected neutron were
reconstructed. Cuts on the latter excluded additional pion production.
Backgrounds from the aluminum target window and random coincidences were
subtracted.  Yields were determined after correcting for tracking 
efficiency, pion absorption, local target-density reduction due to beam 
heating, and dead times. Cross sections were obtained from
the yields using a detailed Monte Carlo (MC) simulation of the experiment,
which included the magnets, apertures, detector geometries, realistic
wire chamber resolutions, multiple scattering in all materials,
reconstruction matrix elements, pion decay, muon
tracking, and internal and external radiative processes.

Calibrations with the overdetermined $^{1}$H$(e,e'p)$ reaction were critical in
several applications. The beam momentum and the spectrometer central 
momenta were determined absolutely to 0.1\%, while the incident beam angle and 
spectrometer central angles were absolutely determined to better than 1 mrad.
The spectrometer acceptances were checked by comparison
of data to MC simulations.
Finally, the overall absolute cross section normalization was checked. The
calculated yields for $e+p$ elastics agreed to better than 2\% with
 predictions based on a parameterization of the world data \cite{bos95}.

In the pion production reaction the experimental acceptances in  $W$, \qsq\ and
$t$ were correlated. In order to minimize errors resulting from averaging the 
measured yields when calculating cross sections at average values of $W$, \qsq\
and $t$, a phenomenological cross section model \cite{thesis} was used in the 
simulation program. 
In this cross section model the terms representing the $\sigma_X$ of 
Eqn.~(\ref{eq:sepsig}) were optimized in an iterative fitting procedure to 
globally follow the $t-$ and $Q^2$-dependence of the data.
The dependence of the cross section on $W$
was assumed to follow the phase space factor $(W^2-M^2_p)^{-2}$.

The experimental cross sections can then be calculated from the measured and
simulated yields via the relation
\begin{equation}
\label{eq:xs}
\left( \frac{d\sigma(\bar W,{\bar Q}^2,t)}{dt} \right)_{\mathrm{exp}} =
\frac{\langle Y_{\mathrm{exp}}\rangle} {\langle Y_{\mathrm{MC}}\rangle}
\left( \frac{d\sigma(\bar W,{\bar Q}^2,t)}{dt} \right)_{\mathrm{MC}}.
\end{equation}
This was done for five bins in $t$ at the four \qsq -values.
Here, $\langle Y\rangle$ 
indicates that the yields were averaged over the $W$ and \qsq\ acceptance, 
$\bar W$ and ${\bar Q}^2$ being the acceptance weighted average values for that
$t$-bin.
Even while the average values of $W$ and $Q^2$ differed slightly at high and
low \eps , the use of Eq.~(\ref{eq:xs}) with a MC cross section that 
globally reproduces the data allows one to take a common average $(\bar W, 
\bar{Q}^2)$ value.

\begin{figure}
    \epsfxsize=3.375in
    \centerline{\epsffile{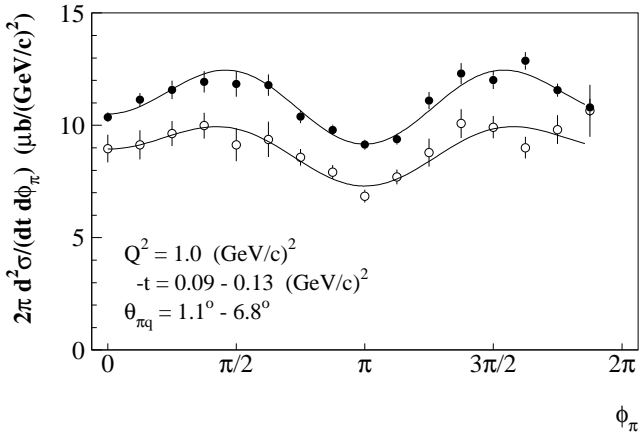}}
\caption{
        $\phi$  dependence of $\frac{d^2 \sigma}{dt d\phi}$ at \qsq =1.0 \gevsq\
        for high and low \eps\ (filled and empty circles, resp.).
        The curves represent the results of the fits.
        \label{fig:phidistr}
}
\end{figure}

A representative example of the cross section as function of $\phi$ is given in
Figure \ref{fig:phidistr}. The dependence on $\phi$ was used to determine the 
interference terms \siglt\ and \sigtt\, after which the combination
$\sigma_{\mathrm{uns}} = \sigma_{\mathrm{T}} + \epsilon \sigma_{\mathrm{L}}$
was obtained at both the high and low electron energy in each $t$ bin for each
\qsq\ point. The statistical uncertainty in these cross sections ranges
from 2 to 5\%. Furthermore, there is a total systematic uncertainty of about 
3\%, the
\noindent most important contributions being: simulation of the detection 
volume  (2\%), dependence of the extracted cross sections on the MC cross 
section model (typically less than 2\%), target density reduction (1\%),
pion absorption (1\%), pion decay (1\%), and the simulation of
radiative processes (1\%) \cite{thesis}.
Since the same acceptances in $W$ and $Q^2$ and the same average values 
$\bar W$ and ${\bar Q}^2$ were used at both energies, \sigl\ and \sigt\
could be extracted via a Rosenbluth separation. 

\begin{figure}[t]
    \epsfxsize=3.375in
    \centerline{\epsffile{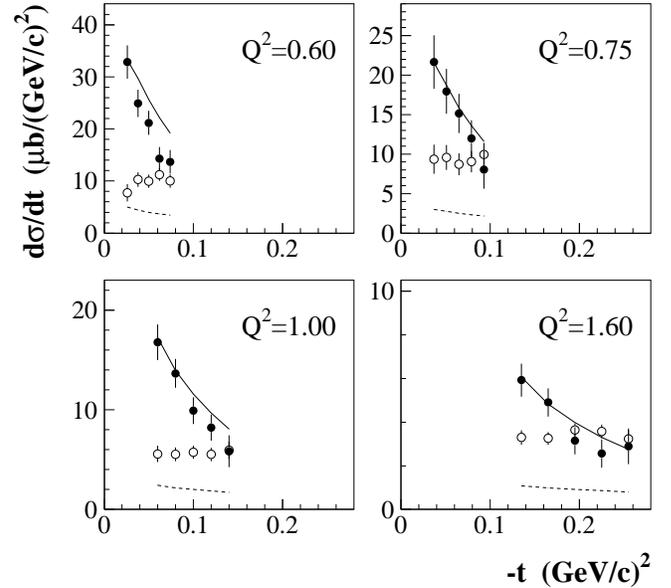}}
\caption{
        Separated cross sections \sigl\ and \sigt\ (full and open symbols,
        resp.) compared to the Regge model (full curve for L,  dashed curve for
        T). The $Q^2$ values are in units of \gevsq .
        \label{fig:cross-sections}
}
\end{figure}

These cross sections are displayed in Figure \ref{fig:cross-sections}. The 
error bars represent the combined statistical and systematic uncertainties.
Since the uncertainties that are uncorrelated in the 
measurements at high and low electron energies are enlarged by the factor 
1/($\Delta \epsilon$) in the Rosenbluth separation, where $\Delta \epsilon$ is 
the difference (typically 0.3) in the photon polarization between the two 
measurements, the total error bars on \sigl\ are typically about 10\%.

The experimental data were compared to the results of a Regge model by 
Vanderhaeghen, Guidal and Laget (VGL) \cite{van97}. In this model the pion 
electroproduction process is described as the exchange of Regge trajectories 
for $\pi$ and $\rho$ like particles. The only free parameters are the pion form
factor and the $\pi\rho\gamma$ transition form factor. The model globally 
agrees with existing pion photo- and electroproduction data at values of $W$ 
above 2 GeV. The VGL model is compared to the data in
Figure \ref{fig:cross-sections}. The value of \fpi\ was adjusted at every \qsq\
to reproduce the \sigl\ data at the lowest value of $t$.  The transverse cross 
section \sigt\ is underestimated, which can possibly be attributed to resonance
contributions at $W=1.95$ GeV that are not included in the Regge model. Varying
the $\pi\rho\gamma$ transition form factor within reasonable bounds changes 
\sigt\ by up to 30\%, but has a negligible influence on \sigl , which is 
completely determined by the $\pi$ trajectory. This t-pole dominance was 
checked by studying the reactions $^2$H$(e,e'\pi^+)nn$ and 
$^2$H$(e,e'\pi^-)pp$, which gave within the uncertainties a ratio of unity for 
the longitudinal cross

\noindent  sections.
Hence the VGL model is still considered to be a good starting point 
for determining \fpi .

\begin{figure}
    \epsfxsize=3.375in
    \centerline{\epsffile{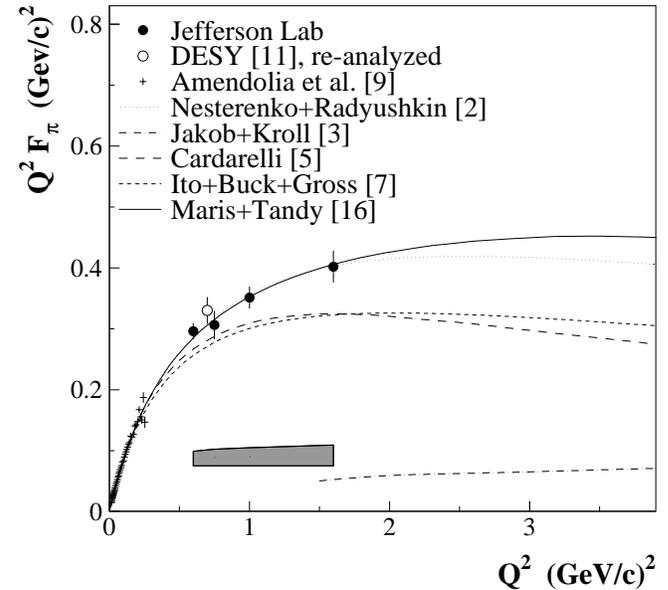}}
\caption{
        The Jefferson Laboratory and reanalyzed DESY values for $F_\pi$
        in comparison to the results of several calculations.
        The model uncertainty is represented by the gray area.
         The (model-independent) data from Ref. \protect\cite{ame86}
        are also shown.
        A monopole behaviour of the form factor obeying the measured
        charge radius is almost identical to the Maris and Tandy curve.
        \label{fig:piff}
}
\end{figure}

The comparison with the \sigl\ data shows that the $t$ dependence in the VGL 
model is less steep than that of the experimental data. 
As suggested by the analysis \cite{gut72} of older data, where a similar
behaviour was observed, we attributed the discrepancy between the data and VGL 
to the presence of a negative background contribution to the longitudinal cross
section, presumably again due to resonances.
Since virtually nothing is known about the effect of these resonances on 
\sigl , 
we proceeded on two paths to determine a trustworthy value of \fpi .
First we fitted the VGL prediction for \sigl\ to the data by adjusting \fpi\ at
the lowest $|t|$ bin, as shown in Fig. \ref{fig:cross-sections}, where it is 
assumed to be most reliable, owing to the dominant $t$ pole behaviour. However,
since there is no reason to believe that the (negative) background is zero at
the lowest $-t$, the result is an underestimate for \fpi .
Secondly, \fpi\ was determined 
adding a \qsq\ dependent negative background to \sigl (VGL) and fitting it 
together with the value of \fpi . The background term was taken to be 
independent of $t$. This was suggested by looking at the 'missing background' 
in \sigt, i.e., the difference between the data and VGL for \sigt. That 
background is almost constant or slightly rising with $|t|$. Then, assuming 
that the background in \sigl\ has a similar $t$-dependence,
a constant background leads to an overestimate of \fpi . Our best
estimate for \fpi\ is taken as the average of the two results. The model 
uncertainty (in relative units) is taken to be the same for the four \qsq\ 
points, and equal to one half of the average of the (relative) differences.
The results are listed in Table \ref{tab:fpi-final} and shown in the form of 
\qsq \fpi\ in Fig.~\ref{fig:piff}. The error bars were propagated from the 
statistical and systematic uncertainties on the cross section data. The model 
uncertainty is displayed as the gray bar. The fact that the value of \fpi\ at 
\qsq\ = 0.6 \gevsq\ is close to the extrapolation of the model independent data
from \cite{ame86}, and that the value of the background term is lower at higher
$W$ (see below), gives some confidence in the procedure used to determine \fpi.

For consistency we have re-analysed the older L/T separated data at \qsq\ = 0.7
\gevsq\ and $W=2.19$ GeV from DESY \cite{bra77}. We took the published cross 
sections and treated them in the same way as ours. The background term in 
\sigl\ was found to be smaller than in the Jefferson Lab data, presumably 
because of the larger value of $W$ of the DESY data, and hence the model 
uncertainty is smaller, too. The resulting best value for \fpi , also shown in 
Fig.~\ref{fig:piff}, is larger by 12\% than the original result, which was 
obtained by using the Born term model by Gutbrod and Kramer \cite{gut72}.
 Here it should be mentioned that those authors used a
phenomenological $t$-dependent function, whereas the Regge model
by itself gives a good description of the $t$-dependence  of the
(unseparated) data from Ref. \cite{beb78}.

The data for \fpi\ in the region of \qsq\ up to 1.6 \gevsq\ globally follow
a monopole form obeying the pion charge radius
\cite{ame86}.
It should be mentioned that the older Bebek 
data in this region suggested lower \fpi\ values.  However, as mentioned, they 
did not use L/T separated cross sections, but took a prescription for 
\sigt . Our measured data for \sigt\ indicate that the values used 
were too high, so that the values for \fpi\ came out systematically low.

In Fig.~\ref{fig:piff} the data are also compared to theoretical
calculations. The model by Maris and Tandy \cite{mar00} provides a good 
description of the data. It is based on the Bethe-Salpeter equation with 
dressed quark and gluon propagators, and includes parameters that were 
determined without the use of \fpi\ data. The data are also well described by
the QCD sum rule plus hard scattering estimate of Ref.~\cite{nes82}. Other
models \cite{car94,ito92} were fitted to the older \fpi\ data and therefore 
underestimate the present data. Figure \ref{fig:piff} also includes the results
from perturbative QCD calculations \cite{jak93}.

In summary, new accurate separated cross sections for the $^1$H$(e,e'\pi^+)n$
reaction have been determined in a kine-
\begin{table}[t]
\caption{
Best values for \fpi\ from the present data and from the
re-analyzed data from Ref. \protect\cite{bra77}.
The total (systematic and statistical) experimental uncertainty is given 
first, and second the model uncertainty.
\label{tab:fpi-final}
}
\begin{tabular}{ccc}
 $Q^2$ \gevsq & $W$ (GeV) & \fpi\ \\
 \hline 
 0.60 & 1.95 & 0.493 $\pm$ 0.022 $\pm$ 0.040 \\
 0.75 & 1.95 & 0.407 $\pm$ 0.031 $\pm$ 0.036 \\ 
 1.00 & 1.95 & 0.351 $\pm$ 0.018 $\pm$ 0.030 \\
 1.60 & 1.95 & 0.251 $\pm$ 0.016 $\pm$ 0.021 \\ \hline
 0.70
       & 2.19 & 0.471 $\pm$ 0.032 $\pm$ 0.037\\ 
\end{tabular}
\end{table}

\noindent matical region where the t-pole process
is dominant. Values for \fpi\ were extracted from the longitudinal
cross section using a recently developed Regge model.
 Since the model does not give a perfect description of
the $t$-dependence of the data, our results for \fpi\ contain a sizeable
model uncertainty. Improvements in the theoretical description of the
$^1$H$(e,e'\pi^+)n$ reaction hopefully will reduce those.
The data globally follow a monopole form obeying the pion charge radius,
and are well above values predicted by pQCD calculations.



The authors would like to thank Drs. Guidal, Laget and Vanderhaeghen for
stimulating discussions and for making their computer program available to us.
This work is supported by DOE and NSF (USA), FOM (Netherlands),
NSERC (Canada), KOSEF (South Korea), and NATO.


\begin{references}
\bibitem{lis92} H.-N. Li and G. Sterman, Nucl.~Phys. {\bf B381} (1992) 129
\bibitem{nes82} V.~A.~Nesterenko and A.~V.~Radyushkin, Phys. Lett.~{\bf B115}
        (1982) 410; A.V. Radyushkin,  Nucl. Phys. {\bf A532} (1991) 141
\bibitem{jak93} R.~Jakob and P.~Kroll, Phys.~Lett. {\bf B315} (1993) 463
\bibitem{bra00} V.M. Braun, A. Khodjamirian and M. Maul,
        Phys.~Rev.~D {\bf 61} (2000) 073004
\bibitem{car94} F.~Cardarelli {\em et al.}, Phys.~Lett.~{\bf B332} (1994) 1;
        Phys.~Lett.~{\bf B357} (1995) 267
\bibitem{ste99} N.G.~Stefanis, W.~Schroers and H.-Ch.~Kim, Phys. Lett.
        {\bf B449} (1999) 299 and hep-ph/0005218
\bibitem{ito92} H.~Ito, W.~W.~Buck and F.~Gross, Phys.~Rev.~C {\bf 45} 
        (1992) 1918
\bibitem{mar98} P.~Maris and C.~D.~Roberts, Phys.~Rev.~C {\bf 58} (1998) 3659
\bibitem{ame86} S.~R.~Amendolia {\em et al.}, Nucl.~Phys. {\bf B277} (1986) 168
\bibitem{beb78} C.~J.~Bebek {\em et al.}, Phys.~Rev.~D {\bf 17} (1978) 1693
\bibitem{bra77} P.~Brauel   {\em et al.}, Z.~Phys. {\bf C3} (1979) 101
\bibitem{thesis} J.~Volmer, PhD thesis, Vrije Universiteit, Amsterdam (2000),
        unpublished
\bibitem{bos95} P.~E.~Bosted, Phys.~Rev.~C {\bf 51} (1995) 409
\bibitem{van97} M.~Vanderhaeghen, M.~Guidal and J.-M.~Laget,
        Phys.~Rev.~C {\bf 57} (1998) 1454; Nucl.~Phys. {\bf A627} (1997) 645
\bibitem{gut72} F.~Gutbrod and G.~Kramer, Nucl.~Phys. {\bf B49} (1972) 461
\bibitem{mar00} P.~Maris and P.~C.~Tandy, preprint nucl-th/0005015

\end{references}
\end{document}